# Magnetospheric Emission from Extrasolar Planets

J. Lazio,[1] Naval Research Laboratory
T. Bastian, NRAO
G. Bryden, Jet Propulsion Laboratory
W. M. Farrell, NASA/GSFC
J.-M. Griessmeier, ASTRON
G. Hallinan, National U. of Ireland
J. Kasper, Smithsonian Astrophysical Observatory
T. Kuiper, Jet Propulsion Laboratory
A. Lecacheux, Observatoire de Paris
W. Majid, Jet Propulsion Laboratory
R. Osten, STScI
E. Shkolnik, Carnegie Institute of Washington
I. Stevens, University of Birmingham
D. Winterhalter, Jet Propulsion Laboratory
P. Zarka, Observatoire de Paris

[1] 202-404-6329; *Joseph.Lazio@nrl.navy.mil*

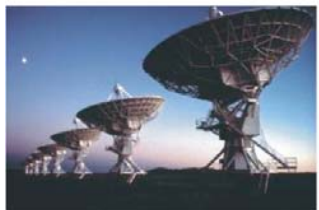
Very Large Array

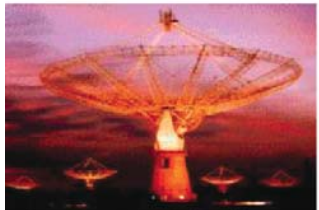
Giant Metrewave Radio Telescope

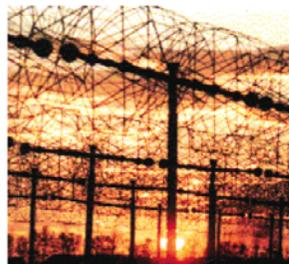
Ukrainian T Radiotelescope

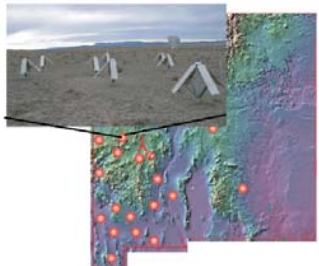
Long Wavelength Array

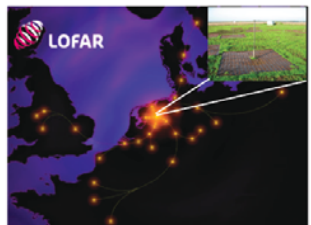
Low Frequency Array

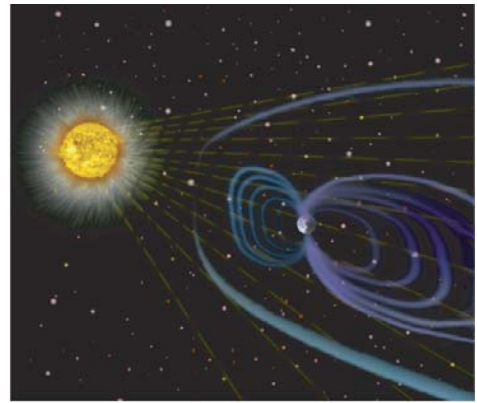

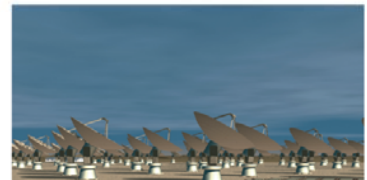
Square Kilometre Array

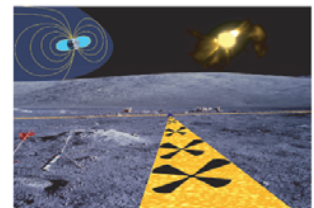
lunar radio array

Image credits: JPL; NRAO; NCRA; IRA; UNM/ONR/NRL; ASTRON; SKA; NASA/GSFC

# Executive Summary

The magnetospheric emissions from extrasolar planets represent a science frontier for the next decade. All of the solar system giant planets and the Earth produce radio emissions as a result of interactions between their magnetic fields and the solar wind. In the case of the Earth, its magnetic field may contribute to its habitability by protecting its atmosphere from solar wind erosion and by preventing energetic particles from reaching its surface. Indirect evidence for at least some extrasolar giant planets also having magnetic fields includes the modulation of emission lines of their host stars phased with the planetary orbits, likely due to interactions between the stellar and planetary magnetic fields. If magnetic fields are a generic property of giant planets, then extrasolar giant planets should emit at radio wavelengths allowing for their *direct* detection. Existing observations place limits comparable to the flux densities expected from the strongest emissions. Additional sensitivity at low radio frequencies coupled with algorithmic improvements likely will enable a new means of detection and characterization of extrasolar planets within the next decade.

# 1 Frontier Question: What is the Luminosity Function of Extrasolar Planetary Magnetospheric Emissions?

Looking to the next decade, the luminosity function of magnetospheric emissions represents a scientific frontier that is ripe for discovery and would present a new means for characterization for extrasolar planets. A combination of algorithmic improvements and developments for existing telescopes as well as the new telescopes under construction should yield at least an order of magnitude of additional sensitivity.

# 2 Science Opportunity: Extrasolar Planetary Magnetospheric Emission

*Search for and exploit extrasolar planetary magnetospheric emissions as a means of directly detecting and characterizing those planets.*

## 2.1 Planetary Magnetospheric Emission

The Earth and gas giants of our solar system are "magnetic planets" because they contain internal dynamo currents that generate planetary-scale magnetic fields. These magnetic fields are immersed in the solar wind, a supersonic magnetized plasma. The solar wind deforms the planetary magnetic field, compressing the field on the front side and elongating it on the back, forming a "tear-dropped"–shaped magnetosphere aligned with the solar wind flow (Figure 1). The magnetopause forms the boundary between the magnetosphere, in which the planet's magnetic field is dominant, and the solar wind. The stellar wind incident on the magnetopause is an energy source to the planetary magnetosphere.

The magnetospheres of the solar system's magnetic planets host radio-wavelength masers, generated by electron cyclotron radiation ("electron cyclotron masers"). The magnetosphere-solar wind interaction produces energetic (keV) electrons that then propagate along magnetic field lines into auroral regions, where an electron cyclotron maser is produced. Specific details of the electron cyclotron maser emission vary from planet to planet, depending upon secondary effects as the planet's magnetic field topology. Nonetheless, applicable to all of the magnetic planets is a macroscopic relation relating the incident solar wind power $P_{sw}$, the planet's magnetic field strength, and the median radio luminosity $L_{rad}$ (Figure 2). Various investigators (e.g.,

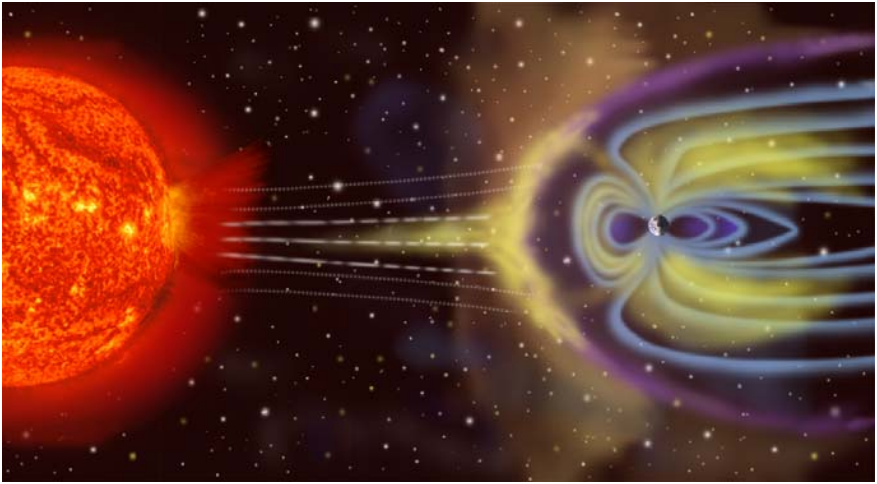

Figure 1. Illustration of the interaction between the Sun and Earth's magnetosphere (*not to scale*). In this example, an eruption from the Sun has reached the Earth, compressed the magnetosphere, and injected energetic particles. Credit: NASA

Millon & Goertz 1988) have found

$$L_{rad} = \varepsilon \, P_{sw}^{x}$$

with ε the efficiency at which the solar wind power is converted to radio luminosity, and $x \approx 1$. The value for ε depends on whether one considers the magnetic energy or kinetic energy carried by the stellar wind. The strong solar wind dependence is manifest in the fact that the Earth's luminosity is larger than that of either Uranus or Neptune, even though their magnetic fields are 10–50 times stronger than that of the Earth.

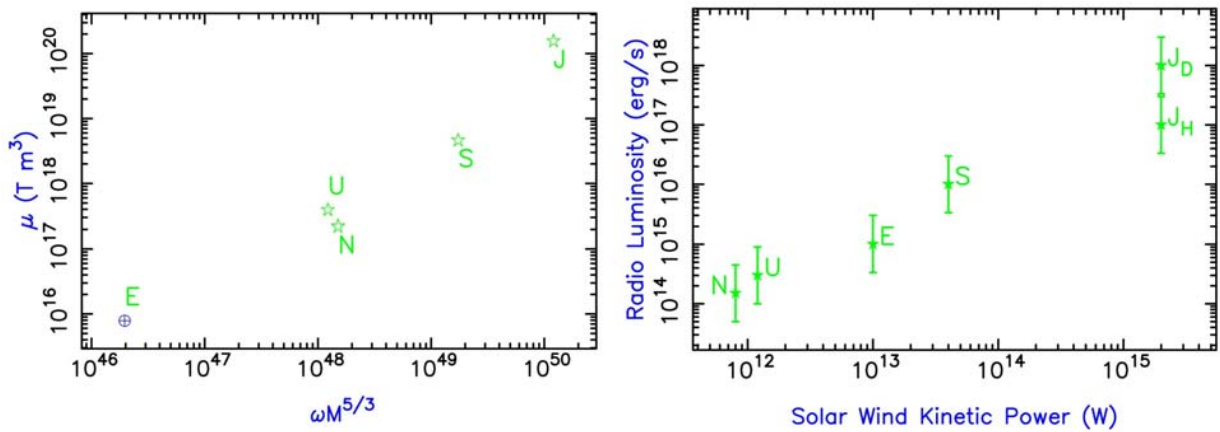

**Figure 2** (*Left*) Magnetic moments of solar system planets as a function of the rotation rate ω and mass M. (*Right*) Radio luminosities of solar system planets as a function of the incident solar wind kinetic power. (Adapted from Zarka et al. 2001.) Even though the magnetic field strength of Earth is less than that of Uranus or Neptune, it is more luminous because it is closer to the Sun and a larger solar wind power is incident. These relations can be both descriptive and predictive, as the luminosity of Uranus and Neptune were both predicted successfully before the arrival of the *Voyager 2* spacecraft.

The electron cyclotron maser occurs below a characteristic emission frequency determined by the cyclotron frequency in the magnetic polar region, which in turn depends upon the planet's magnetic moment or magnetic field strength. Using similar scaling laws based on the solar system planets, one can predict this characteristic frequency. For reference, the characteristic frequency of Jupiter is approximately 40 MHz ($\approx \lambda 7.5$ m).

These scaling relations are not only descriptive but also *predictive*. Before *Voyager 2* reached both Uranus and Neptune, their luminosities were predicted (Desch & Kaiser 1984; Desch 1988; Millon & Goertz 1988). For both planets, the predictions were in excellent agreement with the measurements.

## 2.2 Extrasolar Planetary Magnetic Fields

Indirect evidence for extrasolar planetary magnetic fields is found in modulations of the Ca II H and K lines of the stars HD 179949 and υ And—modulations in phase with planetary orbital periods (Shkolnik et al. 2005, 2008). Photometric observations by the MOST space telescope of several hot-Jupiter systems also suggest that a giant planet can induce stellar surface activity in the form of active spots (Walker et al. 2008).

The solar system scaling laws enable quantitative predictions for an extrasolar planet's luminosity (Zarka et al. 1997; Farrell et al. 1999; Zarka et al. 2001; Lazio et al. 2004; Stevens 2005; Zarka 2007; Griessmeier et al. 2007). In the case of a known planet, these estimates depend upon measured parameters—the planet's mass and orbital semi-major axis—or upon parameters that can be estimated reasonably—for example, the rotation period can be assumed to be of order 10 hr, or for hot Jupiters taken to be the orbital period assuming that the planet is tidally locked to its host star. In the case of stars not yet known to have planetary companions, radio limits can be inverted to obtain constraints on the presence of planets, especially useful for active stars for which the radial velocity method is limited.

Planets around young stars may be more luminous than the solar system planets. Changes in the stellar wind can produce significant changes in a planet's luminosity (Griessmeier et al. 2005a, 2007). Wood et al. (2002, 2005) find the stellar mass loss as a function of age, $dM/dt \propto t^x$, with $x \approx -2$; the stellar wind around a 1 Gyr old star may be 25 times as intense as the solar wind. Young stars are often not observed in radial velocity surveys because the high stellar activity levels make it problematic to isolate a planetary signal. Thus, a blind survey for magnetospheric emissions is a search methodology that could mitigate a selection bias in the current census. As an illustration, early predictions for the flux density of the planet orbiting τ Boo were of order 1–3 mJy (Farrell et al. 1999; Lazio et al. 2004). More recent estimates, that attempt to take into account the stellar wind strength of τ Boo, predict flux densities as high as 300 mJy (Stevens 2005; Griessmeier et al. 2007). If iron-rich "super-Earths" exist, they may also have sufficiently strong magnetic fields to be detectable.

## 3 Scientific Context: Magnetic Fields and Planetary Characterization

The dynamo currents generating a planet's magnetic field arise from differential rotation, convection, compositional dynamics, or a combination of these in the planet's interior. Consequently, knowledge of the planetary magnetic field places constraints on the thermal state, composition, and dynamics of the planetary interior, all of which will be difficult to determine by other means.

**Planetary Interiors:** For the solar system planets, the composition of the conducting fluid ranges from liquid iron in the Earth's core to metallic hydrogen in Jupiter and Saturn to perhaps a salty ocean in Uranus and Neptune. Likewise, radio detection of an extrasolar planet would indicate the planet's internal composition, insofar as it would require the planet to have a conducting interior. Combined with an estimate of the planet's mass and radius, one could infer the interior composition by analogy to the solar system planets.

**Planetary rotation:** The rotation of a planet imposes a periodic modulation on the radio emission, as the emission is preferentially beamed in the direction of the local magnetic field and will change if the magnetic and spin axes of the planet are not aligned. For all of the gas giant planets in the solar system, this modulation *defines* the rotation periods. For instance, the rotation period of Neptune was determined initially by observations of differentially rotating cloud tops but then was redefined after detection of its radio emission (Lecacheux et al. 1993).

**Planetary Satellites:** In addition to being modulated by its rotation, Jupiter's radio emission is affected by the presence of its satellite Io, and more weakly by Callisto and Ganymede. As the Jovian magnetic field sweeps over a moon, a potential is established across the moon by its **v** × **B** motion in the Jovian magnetic field. This potential drives currents along the magnetic field lines, connecting the moon to the Jovian polar regions, where the currents modulate the radio emission. Modulations of planetary radio emission may thus reveal the presence of a satellite.

**Atmospheric retention:** A common and simple means of estimating whether a planet can retain its atmosphere is to compare the thermal velocity of atmospheric molecules with the planet's escape velocity. If the thermal velocity is a substantial fraction of the escape velocity, the planet will lose its atmosphere. For a planet immersed in a stellar wind, *nonthermal* atmospheric loss mechanisms can be important (Shizgal & Arkos 1996), as the typical stellar wind particle has a *supra-thermal* velocity. If directly exposed to a stellar wind, a planet's atmosphere can erode more quickly. Based on Mars Global Surveyor observations, this erosion process is thought to have been important for Mars' atmosphere and oceans (Lundin et al. 2001; Crider et al. 2005).

**Habitability:** A magnetic field may determine the habitability of a planet by deflecting cosmic rays or stellar wind particles, (e.g., Griessmeier et al. 2005b). In addition to its effect on the atmosphere, if the cosmic ray flux at the surface of an otherwise habitable planet is too large, it could cause cellular damage or frustrate the origin of life altogether.

## 4  Science Requirements and Key Advances for the Next Decade

*The exploitation of extrasolar planetary magnetospheric emissions will require sensitive, low radio frequency observations.*

Table 1 summarizes key requirements for detecting and exploiting extrasolar planetary magnetospheric emissions. In the rest of this section we motivate these requirements.

**Table 1.  Scientific Requirements for Extrasolar Planetary Magnetospheric Observations**

| Parameter | Value | Comment |
|---|---|---|
| Frequency (Wavelength) | <~ 100 MHz (>~ 3 m) | Determined by planetary magnetic field; Brown dwarf observations suggest higher frequencies possible. |
| Sensitivity | < 25 mJy | Extrapolations from solar system relations; Constrained by existing observations. |

The characteristic frequency above which magnetospheric emissions are no longer generated is determined by the strength of the planet's magnetic field; Jupiter's magnetic field strength, at the cloud tops, is approximately 4 G, leading to a characteristic (cutoff) frequency of approximately 40 MHz. Allowing for the known extrasolar planets more massive than Jupiter, observations below 100 MHz (> 3 m) likely will be required.

The sensitivity requirement is motivated both by extrapolations of magnetospheric emissions from solar system planets as well as current observational limits (§4.1). Extrapolations from solar system planets suggest that at lower frequencies, magnetospheric emissions are stronger so that a somewhat less stringent requirement suffices. Also, because the emissions are driven by the stellar wind, planets closer to their host stars are likely to be brighter.

The performance of a low frequency radio telescope is determined by a combination of design parameters (notably its collecting area), the calibration of the telescope, and the identification and excision of radio frequency interference (RFI).

For low radio frequency observations of extrasolar planets, improved sensitivity generally requires increasing the effective collecting area of the telescope $A_{eff}$. The theoretical sensitivity of a radio telescope, as given by the radiometer equation, is determined by $A_{eff}$, its system temperature $T_{sys}$, the observation bandwidth, and integration time. The magnetospheric emission from solar system planets is broadband, and modern radio astronomical systems process bandwidths that are a large fraction of the observing frequency. Below 100 MHz, the Galactic nonthermal emission makes a significant contribution to the system temperature $T_{sys}$. Finally, magnetospheric emission from solar system planets can be "bursty," so that long integrations may not be as useful as for traditional observations, because a long integration may average together times when the planet is emitting with times when it is not.

Phase distortions induced by the Earth's ionosphere also impact the sensitivity of low frequency radio telescopes. In the past decade, new calibration schemes have been developed for the ionospheric phase corruptions at radio wavelengths, often bearing a conceptual similarity to adaptive optics techniques for tropospheric phase corruptions at visible wavelengths. Modern low frequency radio telescopes plan both to exploit these new algorithms as well as develop new algorithms in order to reach the theoretical noise limit implied by the radiometer equation.

The final impact on performance is RFI produced by civil or military transmitters operating at similar frequencies. These transmitters are often orders of magnitude stronger than the desired signal. Current RFI identification and excision techniques are relatively crude. More sophisticated algorithms are among the key development aspects for the next generation of radio telescopes.

## 4.1 Current Status

Magnetospheric emissions from the solar system planets motivated searches for analogous emissions prior to the discovery of extrasolar planets (Yantis et al. 1977; Winglee et al. 1986). Two significant changes in the past decade have been the discovery of extrasolar planets and the development of low frequency telescopes with an order of magnitude more sensitivity than historical telescopes.

There have been multiple searches toward a number of the known extrasolar planets, with many of these searches still in progress (Table 2). The premier instruments are the Very Large Array (VLA) with its 74 MHz system, the Giant Metrewave Radio Telescope (GMRT) with its 150 MHz system, and the Ukranian T-shaped Radiotelescope (UTR-2), which observes at 8–31 MHz. The UTR-2 observes at low enough frequencies that it can detect Jovian emissions, and the 74-MHz VLA observes at a frequency that is within a factor of two of the highest frequency Jovian emissions (40 MHz).

Table 2. Limits on Extrasolar Planetary Magnetosphere Emission[2]

| Frequency | Limit | Telescope | Reference |
|---|---|---|---|
| 150 MHz | 0.3–2 mJy | GMRT | Hallinan et al. 2009; Winterhalter et al. 2009 |
| 74 MHz | 135–300 mJy | VLA | Lazio & Farrell 2008 |
| 25 MHz | 100–1600 mJy[3] | UTR-2 | Zarka 2007 |

The most effort has been directed toward the planet orbiting τ Boo. This planet's luminosity is $L < 10^{23}$ erg s$^{-1}$, unless its radiation is highly beamed into a solid angle $\Omega \ll 1$ sr, much smaller than that of any of the solar system planets. This luminosity limit is lower than some, but not all, recent predictions. Although higher sensitivity observations are likely required, the non-detection may also be hinting that the magnetic fields, and internal compositions, of extrasolar planets are as varied as the planets themselves.

### 4.2  Research and Analysis Advances for the Next Decade

There are a series of algorithmic and theoretical developments that could lead to the detection and exploitation of extrasolar planetary magnetospheric emissions with existing telescopes:

- **Key algorithm development** includes more sophisticated RFI excision and time-frequency processing of the signals, particularly, pattern-matching algorithms that could use Jovian radio emission as a template for higher significance detections.
- **Continued polarimetric observations of host stars** to characterize phenomena related to star-planet interactions (§1.2). These observations will allow fundamental planetary characteristics to be extracted, particularly when combined with modeling efforts (e.g., Lanza 2008; Kitiashvili & Gusev 2008).
- **Development of algorithms that may be deployed in firmware** to carry out active adaptive nulling of bright sources in antenna primary beams. Such techniques are currently computationally prohibitively expensive, but may be feasible with FPGA or GPU technologies.
- **Support for coordinated multi-wavelength observations with space missions**—some preliminary work has already been done with COROT and GMRT observations.

### 4.3  New Telescopes in the Next Decade

In construction are the Long Wavelength Array (LWA, New Mexico) and the Low Frequency Array (LOFAR, the Netherlands). The LWA will operate in the 20–80 MHz band; LOFAR will operate in the 30–80 MHz and 110–240 MHz bands. Both instruments cover the frequency range expected for emission from Jovian-mass to several Jovian mass planets, with sensitivities expected to be 1–2 orders of magnitude better than existing facilities. LOFAR's initial operational phase is anticipated to be late 2009, with the LWA following within a few years.

Looking to the latter half of the next decade and beyond, the Square Kilometre Array (SKA) is a next-generation telescope in the design and development phase that is expected to operate above

---

[2] We list telescopes rather than planets targeted, because multiple searches have been conducted, including the same planet at more than one telescope.

[3] The UTR-2 data acquisition system has been upgraded. The quoted limits are from the old system, as the analysis of current extrasolar planet searches is still in progress.

70 MHz. Its design goals are such that it should be easily capable of detecting the radio emissions from the most massive extrasolar planets. However, a significant constraint to all ground-based telescopes is the Earth's ionosphere, which is opaque below about 10 MHz. The Earth's magnetosphere emits auroral kilometric radiation (AKR) below 1 MHz. Thus, the detection of AKR from extrasolar terrestrial-mass planets—and assessments of their habitability—can only be accomplished from space. The most promising location for a telescope designed to detect AKR from extrasolar terrestrial-mass planets is the far side of the Moon, as it would always be shielded from the Earth.